\documentclass[prb,aps,floatfix,preprint]{revtex4-1}

\usepackage[final]{graphicx}
\usepackage[usenames]{color}
\usepackage{afterpage}
\usepackage{amssymb}
\usepackage{textcomp}

\begin{document}

\title{Theoretical investigation of FeTe magnetic ordering under hydrostatic pressure}
\author{ M. Monni$^{1,2}$, F. Bernardini$^{1,2}$,  G. Profeta$^{3,4}$ and S. Massidda$^{1,2}$}
\affiliation{1 Dipartimento di Scienze Fisiche, 
Universit\`a degli Studi di Cagliari, Cittadella Universitaria, I-09042 Monserrato (CA), Italy}
\affiliation{2 CNR-IOM (UOS Cagliari), Cittadella Universitaria, I-09042 Monserrato (CA), Italy}
\affiliation{3 Dipartimento di Scienze Fisiche e Chimiche,
Universit\`a degli Studi dell'Aquila, Via Vetoio 10,
I-67100 Coppito, L'Aquila, Italy}
\affiliation{4 CNR-SPIN, Via Vetoio n. 10, I-67100 Coppito, L'Aquila, Italy}

\begin{abstract}

We investigate the pressure phase diagram of FeTe, predicting  structural and  magnetic properties in the normal state
at zero temperature within density functional theory (DFT).
 We carefully examined several possible different crystal structures over a pressure range up to 
 $\approx 30 $ GPa:
simple tetragonal (PbO type), simple monoclinic, orthorhombic (MnP type),
 hexagonal (NiAs and wurzite type) and cubic (CsCl and NaCl type).
We predict  pressure to drive
the system through different magnetic ordering (notably also some ferromagnetic phases) eventually 
suppressing magnetism at around 17GPa. 
We speculate the ferromagnetic order to be the reason for the absence of a superconducting phase in FeTe at variance with the case of FeSe.
\end{abstract}
\maketitle

\section{introduction}
In the chalcogenide family of iron-based superconductors\cite{review-takano}
(FeTe$_{1-x}$Se$_{x}$, with $0\leqslant x\leqslant 1$) the most studied are $x=1$ (FeSe) and alloys near the optimal doping
$x\simeq0.5$ (FeTe$_{0.5}$Se$_{0.5}$). The latter has till now the highest superconducting critical
 temperature $T_c=15.6$K\cite{Tropeano2010} among the chalcogenides at zero pressure, while the first shows a $T_c$ which depends strongly on
external pressure: $T_c$ increases from 8K at ambient pressure up to 37K at $P\sim$9 GPa.
\cite{Margadonna2009,Medvedev2009}
On the other hand, FeTe  ($x=0$) is a noticeable example of non-superconducting parent compounds, 
in spite of having peculiar magnetic properties with potentially better superconducting properties than FeSe\cite{Zhang2009}.

However, pure FeTe is always synthesized in the non stoichiometric form Fe$_{1+y}$Te.\cite{Zhang09,Marina10} 
At higher concentrations it is also found that the
 excess Fe plays a role in determining the magnetic properties.\cite{Zhang2009,Stock2011}

The ground state of FeTe is experimentally found as double stripe antiferromagnetically ordered phase (AFMs2)\cite{Bao2009} and theoretically confirmed.\cite{Ma2009,Moon2010,Profeta2012}
 The AFMs2 ordering consists into an AFM  alternation of pairs of ferromagnetically ordered stripes of Fe-atoms,
 and can be seen as a spin-density wave (SDW) with a wave vector half of that corresponding to the usual
 stripe AFM ordering found in pnictides.
This magnetic phase survives at low temperature with no sign of superconducting phase transition. This finding stimulated the search of a possible superconducting phase of FeTe, in particular, looking for a way to destroy the antiferromagnetic phase thus enhancing the spin-fluctuations.

For this purpose, hydrostatic pressure (P) has been  largely used to induce superconductivity in non-superconducting materials at ambient pressure. 
In Fe-based superconducting compounds (both  pnictides and chalcogenides), this technique had many successes
\cite{Zhang09,Okada2009,Giannini2010,Duncan,Kreyssig2008,Kimber,Mittal,Uhoya,Colombier,Torikachvili2008,Yamazaki,
Goldman,Pratt,Mani}, guiding the discovery of new superconducting materials.

Regarding FeTe, transport measurements\cite{Okada2009} show that pure FeTe is not a superconductor even under high pressure ($P$ up to $\sim 19$GPa). The same paper shows the presence of several phase transitions on FeTe
by varying  temperature and pressure (see Fig. \ref{fig:P_V1}).

Zhang \emph{et al.}\cite{Zhang09} observe a sudden decrease of the isothermal compressibility
 for $P\gtrsim3$GPa.
They observe that for $P\lesssim3$ GPa the $c$ axis decreases with increasing pressure
 much less than for $P\gtrsim3$ GPa.
The $a$ axis, on the other hand, decreases with the same slope
with increasing pressure in the whole investigated range ($P$ up to $\approx$10 GPa).
This sudden kink in the slope of the $c$ axis versus $P$ curve was supposed to be related to some phase
 transitions
 (possibly of magnetic origin) and was observed to be second order (without any discontinuity in the
 unit cell volume as a function of pressure). 

Another paper\cite{Giannini2010} reports isothermal compressibility data for $P$ less than 10GPa: it is quite smooth and does not show any derivative
 discontinuity.

 However, these experimental evidences are at variance with what observed in the case of 122 compounds 
(Ca(Ba)Fe$_{2}$As$_{2}$ for example), in which a first order phase transition is observed\cite{Torikachvili2008} with volume contraction, well predicted by first principle calculations.\cite{Yildirim2009,Ji2011,Colonna2011}
In the 122 type compounds a symmetry preserving  phase transition, driven by  hybridization of the p$_z$ As orbitals,
produces a sudden shortening of the $c$-axis
and a sudden transition to a compressed phase (with a discontinuity in the volume versus pressure
 curve).\cite{Colonna2011} 
 In FeTe, on the other hand, the crystal structure does not allow direct Te-Te bonds along $z$-direction, thus this same phase transition is not expected at all.

Recently, by means of muon spin rotation, dc magnetization, and neutron depolarization measurements\cite{Bendele2012} new magnetic phases were reported (never reported for 1111 and 122 class of superconductors) 
and with synchrotron powder x-ray diffraction\cite{Koz2012} a plethora of structural 
phase transitions were observed within the 0-3 GPa pressure range in Fe$_{1.08}$Te.

Within this scenario it is evident that the structural phase diagram of FeTe under pressure is not complete and the experimental evidences are not deeply explored and are  still unclear, 
even from a theoretical point of view. 
In particular there are no informations on the possible structural phases under pressure, on new magnetic orderings and on the theoretical possibility to  obtain an high-pressure non-magnetic phase.
 
In order to clarify the role of pressure in determining the interplay between structural and magnetic properties of FeTe, we 
performed extended first-principles simulations of the high-pressure phase diagram of FeTe within the Density Functional Theory (DFT) considering many different phases, both in their non-magnetic and magnetic configurations.

\section{computational method}
Our calculations were performed within DFT 
 framework  using  the Perdew-Wang\cite{Perdew1992} version of generalized gradient approximation (GGA).
The Projected Augmented Waves (PAW) pseudopotentials\cite{Blochl} were used for both  Fe and Te atom as  implemented in the VASP package.\cite{Kresse1,Kresse2}. 
Since usually Fe 3p semi-core electrons are treated as valence electrons in other Fe-based superconductors\cite{Mazin2008} and in order to ensure a proper transferability in the high pressure region (see below) we included them as valence electron 
too.
Most of calculations were performed using a super cell with 8 Fe atoms and 8 Te atoms (see also below), in order to
 deal with the zero pressure equilibrium state magnetic ordering.
We use a plane waves cutoff of 500eV. The k point mesh was a $6\times6\times6$ Monkhorst Pack grid for the
tetragonal PbO structure ($2a\times 2a\times c$ unit cell, see below), in other cases a comparable mesh
 was chosen.

\begin{figure}
\includegraphics[scale=0.35]{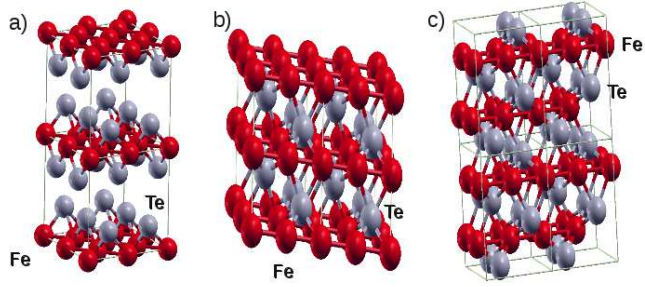}
\caption{(Color online). Panel (a) PbO structure. Panel (b) NiAs structure. Panel (c) MnP structure. The number of atoms
was chosen for the sake of clarity and it is not related to that used in the simulations.}
\label{fig:strutture}
\end{figure}

We consider several possible different crystal structures (see Fig.\ref{fig:strutture}):
simple tetragonal (PbO type, $P4/nmm$ space group), monoclinic, orthorombic (MnP type), hexagonal (NiAs and wurzite type)
 and cubic (CsCl and NaCl type). All structures were simulated with 8 Fe atoms unit cells. 
 For tetragonal PbO cell we use a $2a\times 2a\times c$ unit cell ($a$ and $c$
 referred to the 2Fe atoms unit cell). The tetragonal phase is  unstable when an antiferromagnetic order is 
 imposed and lowers the symmetry into  
 an orthorhombic or  monoclinic phase, slightly distorting the Fe in plane squares (AFM phase) or the angle between $a$ and $c$ (AFMs2 phase).
The MnP type phase is an orthorhombic phase of the $Pbnm$ space group and can be viewed as a lower symmetry
 distortion of an hexagonal NiAs phase (space group $P6_{3}nmc$).
The distortion of the NiAs phase to the MnP phase is customary in this family of compounds, as discussed for
FeSe.\cite{Margadonna2009,Naghavi2011}

We choose a set of volumes spanning the interval 40\AA$^3$/2Fe -- 108\AA$^3$/2Fe
and for each one we find the minimum energy configuration, at fixed volume, relaxing the internal atomic positions and cell parameters.
In order to determine  the transition pressures, we calculate the enthalpy $H=E+PV$ as a function of the pressure.
The pressure $P$ was calculated as the trace of the stress tensor (always isotropic). We also calculated the pressure
from the analytic derivative of the $E(V)$ curve obtained fitting the $E(V)$ data with a Birch-Murnaghan
equation of state. The two methods are in good agreement as far as the Birch-Murnaghan equation of state fits the calculated total enegies.

We investigate various magnetic orderings: non magnetic (NM), collinear antiferromagnetic stripe
 (AFMs), antiferromagnetic double stripe (AFMs2, also called bicollinear in the literature)\cite{Ma2009,Moon2010},
 antiferromagnetic checkerboard (CB) and ferromagnetic (FM).
Spin-polarized calculation were performed in the collinear approximation without including spin-orbit coupling.

\section{results and discussion}
\label{resul}

\begin{figure}
\includegraphics[scale=0.35]{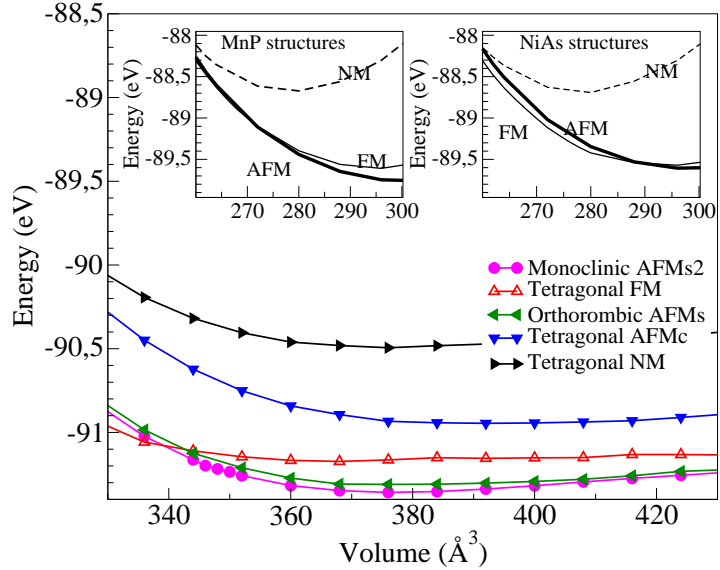}
\caption{(Color online). Total energy versus unit cell volume for an 8 Fe atoms cell. 
For clarity, two different pressure and energy ranges are reported in different panels. }

\label{fig:etot_2panel}
\end{figure}

Fig. \ref{fig:etot_2panel} shows the total energy as a function of unit cell volume for an 8 Fe atoms unit cell.
The most stable structure at $P=0$ is the monoclinic crystal structure  with an AFMs2 magnetic ordering.
 The distortion with respect to tetragonal symmetry is small and driven by the magnetic pattern, which brakes the equivalence of $a$ and $b$ axis.
This can be rationalized in terms of a frustrated Heisenberg model.\cite{Ma2009}

The $a$ and $b$ axis are no longer equivalent as in the tetragonal structure and the angle between $a$ and $c$
 axis is no longer 90\textdegree\ but ranges from 88.35\textdegree\ and 88.2\textdegree\
for $P\leq$2GPa, in a suitable agreement with what is found  experimentally (89.17\textdegree) and in line with other theoretical calculations.\cite{Marina10, Profeta2012}
The atoms internal coordinates vary (with respect to those in the tetragonal structure) in such a way that
  stripes with the same spin orientation become closer and  stripes with opposite spin orientation
 become more distant. 

 The AFMs2 phase has an energy gain of less than 6 meV/Fe with respect to AFMs phase,
which is the ground state of most of other Fe-based superconductors. 
This theoretical prediction agrees with both experiments\cite{Li2009} and theoretical calculations.\cite{Ma2009,Moon2010}
At $P=0$ the equilibrium volume is $V_{0}$=94\AA$^{3}$/2Fe, in good agreement with experimental
 values reported in literature: 93.2\AA$^{3}$/2Fe at ambient temperature\cite{Zhang09},
92.96\AA$^{3}$/2Fe at T$\leq$45K \cite{Kuo-Wei}, 90.78\AA$^{3}$/2Fe at T$\leq$2K\cite{Marina10}.
The slightly overestimation of the calculated volume with respect to low temperature measurements is usual
 in some material for GGA functional and in this case is  mainly due to an overestimation of the out-of-plane ($c$) lattice constant.

With respect to the  AFMs2 phase, the other tetragonal phases have  energies higher by
 $\sim$6 meV/Fe (AFMs), $\sim$24 meV/Fe (FM), $\sim$53 meV/Fe (CB), and $\sim$108 meV/Fe (NM)
 at $P=0$. These energy differences appear to be
 qualitatively in agreement with published theoretical results\cite{Ma2009,Moon2010}, which in turn differ
 from each other on the same energy scale (see Tab.~\ref{tab:deltaE}). 
 Note that both papers\cite{Ma2009,Moon2010} use different experimental lattice constants, while we optimized them.

At volumes near the equilibrium one, for $V\geqslant340$\AA$^3$/8Fe ($V\geqslant85$\AA$^3$/2Fe),
 the low energy phases are the tetragonal derived ones (slightly distorted depending on the magnetic order);
 at lower volumes ($V\leqslant340$\AA$^3$/8Fe), on the opposite,
the low energy phases become the NiAs derived ones (NiAs and MnP). 
All the other phases considered (CsCl, NaCl
 and wurzite) lie at higher energies at all considered volumes irrespective of their magnetic ordering (we do not
 show them in the figures for clarity, lower panel, Fig.\ref{fig:etot_2panel}).

%

\begin{widetext}
\begin{center}
 \begin{table*}[h!]
 \hfill{}
    \begin{tabular}[t]{|l|c|c|c|}
\hline
      Mag. phase    &  This work & Moon \textit{et al}.                & Ma \textit{et al.} \\
\hline
       AFMs2        &    0       &   0                        &   0 \\
\hline
       AFMs         &    6       &  30                        &  10 \\
\hline
       FM           &   24       &  70                        &  76 \\
\hline
       CB           &   53       & 125                        &  68 \\
\hline
       NM           &  108       & n.d.                       & 166 \\
\hline
    \end{tabular}
\hfill{}
\caption{Total energy of different magnetic orderings at zero pressure (meV) with respect to the AFMs2 ground state.
 The results are calculated in the tetragonal structure for the sake of comparison.}
\label{tab:deltaE}
\end{table*}
\end{center}
\end{widetext}

It is worth mentioning that even if  low pressure distortions are sizable they do not determine the relative energy ordering 
 of the different magnetic phases.
The ground state is still AFMs2 and the first transition is still towards a ferromagnetic
 ordered phase even fixing the crystal structure in the tetragonal phase (see below). This is clear looking at Fig.~\ref{fig:etot_tetragonal}, were we report the total
 energy of the different magnetic structures  without (in the tetragonal phase) the magnetically-induced structural
 distortions.

\begin{figure}
\includegraphics[scale=0.35]{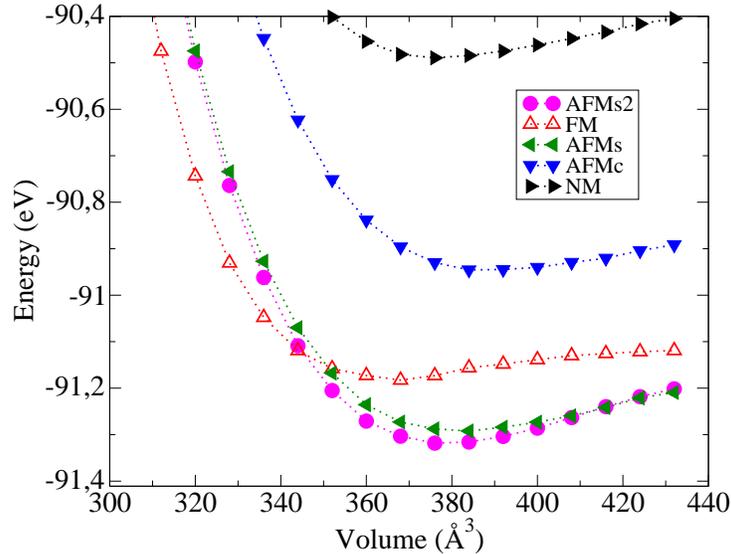}
\caption{(Color online). Total energy (8 Fe atoms unit cell) versus volume for the most relevant magnetic orderings
in the undistorted tetragonal cell.}
\label{fig:etot_tetragonal}
\end{figure}

\begin{figure}
\includegraphics[scale=0.5]{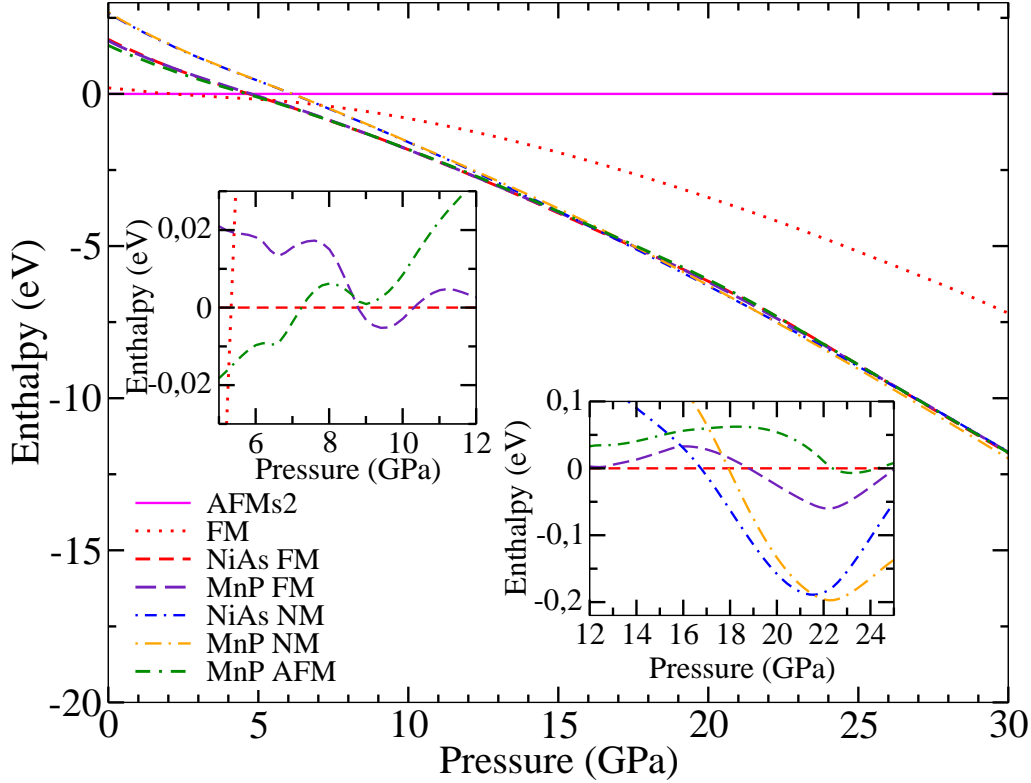}
\caption{(Color online). Enthalpy $H$ as a function of pressure. In the main panel $H$ is relative to that
 of the zero pressure stable phase (AFMs2), while in the insets it is relative to that of the NiAs FM phase.}
\label{fig:P_H1}
\end{figure}

Having discussed the ground state geometry, at P=0, we pass to discuss  high-pressure region of the phase diagram.
Fig.~\ref{fig:P_H1} shows the enthalpy  of FeTe over a wide range of pressures. It is evident how 
the system goes through
 several phase transitions going from lower to higher pressures. This is in line with what was recently observed with x-ray diffraction\cite{Koz2012}.
Starting from the monoclinic AFMs2, a first transition leads to a tetragonal FM structure at 2.1 GPa. 
In this phase Fe layers are surprisingly ferromagnetically ordered. 
Around 5.3 GPa there is a further transition. Increasing  $P$ we find many phases which compete with each other and with very similar enthalpies.
Warning the reader that our predictive power on this scale
may be limited by, \emph{e.g.}, the accuracy in the computed pressure, the sequence of phases encountered is:
MnP AFM (at P= 5.3 GPa), NiAs FM (at P= 7.2 GPa), MnP FM (at P=8.8 GPa), NiAs FM (at P=10.0 GPa), NiAs NM (at P=17 GPa), MnP NM (at P=22 GPa).
The system eventually becomes non-magnetic at $P=$17 GPa. 
We point the experimental attention in this pressure range, in order to discover possible non-magnetic phases. 
However, the crystal symmetry and the magnetic alignment (FM) of the 
closer  phase (ferromagnetic NiAs phase) does not favor a superconducting phase, at least assuming 
an antiferromagnetic spin-fluctuation mechanism.

\begin{figure}
\includegraphics[scale=0.5]{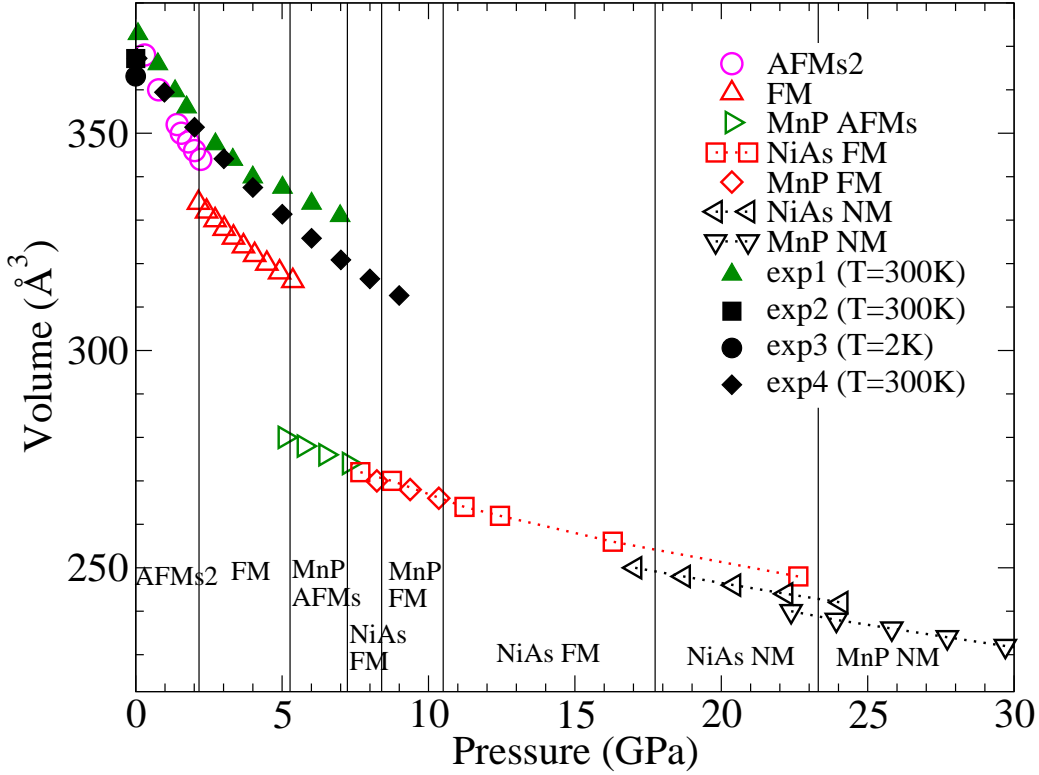}
\caption{(Color online). Cell volume as a function of pressure. The experimental curves (full black symbols) are taken from Ref. \onlinecite{Zhang09} (exp1), from Ref. \onlinecite{Marina10} (exp2 and exp3) and from Ref. \onlinecite{Giannini2010} (exp4).}
\label{fig:P_V1}
\end{figure}

Fig.~\ref{fig:P_V1} shows the volume of an 8 Fe atoms cell as a function of pressure.
The experimental equilibrium volume is well reproduced, while the bulk modulus $B_0$ is underestimated
 (i.e. the compressibility is overestimated) with respect to experiment. The fit with a $3^{rd}$ order
 Birch-Murnaghan equation of state
gives $B_0$=13.1GPa, while experimentally a value of $B_0$=31.3GPa at 300K was reported\cite{Zhang09}.
The  $P=$2.1GPa transition to the FM tetragonal structure is of first order   and  
is accompanied by a volume reduction of $\sim$2.5\AA$^3$/2Fe.

At $P\simeq$1.5 GPa, Okada \emph{et al.}\onlinecite{Okada2009} indicates the presence of a phase transition by
means of magnetic and transport measurements taken at 4.2 K. It is argued
that the monoclinic antiferromagnetic phase (low temperature and low pressure) is suppressed by increasing
 pressure in favor of another phase with a different magnetic state. According to our results this phase could be the
 FM tetragonal structure.
 This phase transition is very peculiar, given the apparent similarity of FeTe with other members of the iron-based pnictides compounds. However there are strong experimental indications. In fact, 
 recently Bendele \emph{et al.}\cite{Bendele2012} report on ferromagnetism induced by pressure in Fe$_{1.03}$Te. They
argue that at low temperature there is a transition from AFM to FM ordering, which confirms our predictions.
Structural measurements\cite{Koz2012} indicate that Fe$_{1.08}$Te at low temperature and high  pressure
undergoes a structural transition from a monoclinic structure to a tetragonal structure at a pressure of about 1.5GPa. 
Even this structural phase transition is nicely predicted by our calculations which predicts a tetragonal FM phase at 2.1 GPa.

We point out that although a small amount of excess Fe in samples may have a role in driving the system properties,
eventually producing an apparent agreement between experiments and calculations, our results seem to rule out also this possibility (see below for calculations on the Fe excess role). 

Other experiments\cite{Zhang09} claim a second order transition at
$P\sim$3 GPa which brings the system in the so called ``compressed tetragonal'' phase, with the caveat that
 those data were taken at 300 K. Our results do not seem to support  this conclusion (see later and discussion in the Introduction).
For $P$ in the interval $(2,19)$GPa Ref.\onlinecite{Okada2009} does not show evidence of other phase transitions, apart
 from one at $P\approx 10$ GPa, suggested on the basis of the electrical resistance versus temperature curves.
 These show a  qualitatively different behavior at $P$ above and below $10$ GPa, but the authors declare that this could
 likely be due to a non hydrostatic stress.
 
Increasing the pressure a $1^{st}$ order transition from tetragonal FM towards another AFM phase (in the MnP structure) takes place at $P=$5.3 GPa, with a sizable volume change. For $P$ going from 5.3GPa to 17GPa the system undergoes
some transitions between MnP and NiAs magnetic phases (the MnP phase is $de-facto$ a distorted NiAs phase).
 These phases are almost degenerate because in this pressure range the distortion that brings the NiAs
 to an MnP phase is very small.
 At $P=$17GPa the magnetization goes definitely
 to zero and the system adopts a non magnetic NiAs structure, which distorts to a non-magnetic MnP structure
 at $P=$22GPa. For higher pressures we do not find any other phase transitions, the MnP phase becomes more and
more distorted with respect to NiAs one. The distortion is essentially characterized by  a buckling of the Fe planes, which are
no longer flat.  This last buckling of the Fe plane has interesting effects even at zero pressure: in fact the 
AFMs2 magnetic phase is easily destroyed by the Fe buckling.

\begin{figure}
\includegraphics[scale=0.5]{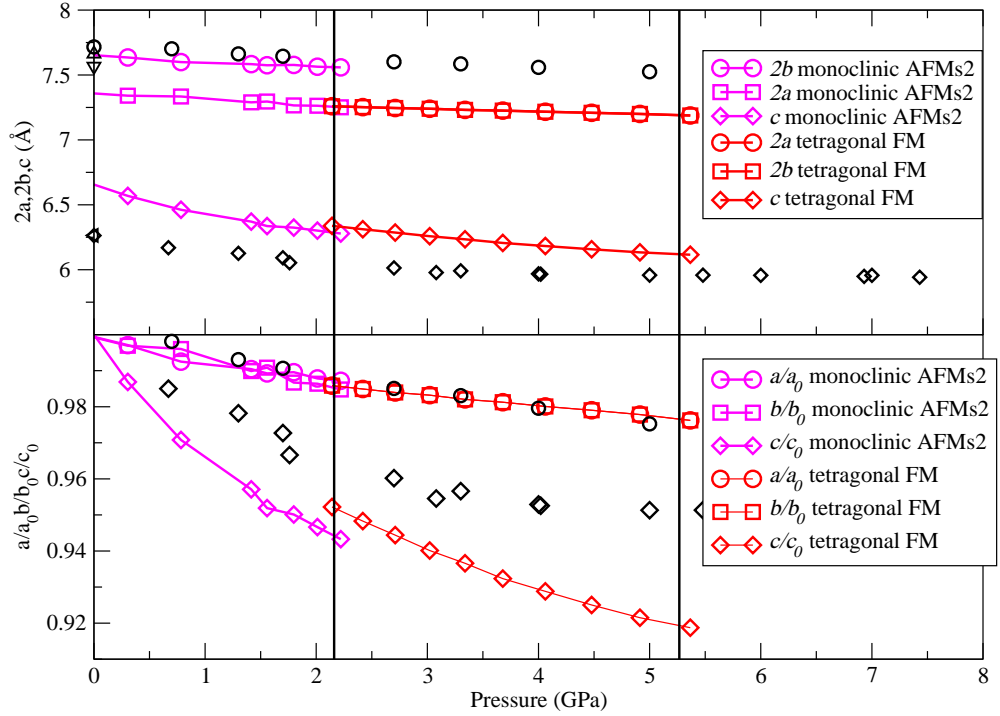}
\caption{(Color online). Cell parameters at low pressure compared to experimental results. Top-panel: absolute value.
 Bottom-panel: values relative to those at $P=$0 as function of pressure.
 Experimental data from Ref.~\onlinecite{Zhang09,notazhang} refined in the PbO ($P4/nmm$) structure:
 circles and rhombus; data from Ref.~\onlinecite{Marina10} refined in the $Pbnm$ structure
 ($P2_1/m$ space group): triangles (only zero pressure).}
\label{fig:Pvasp_csuc0_asua0}
\end{figure}

Fig. \ref{fig:Pvasp_csuc0_asua0} shows low pressure cell parameters, both in absolute units and relative to
 those at $P=$0GPa compared to experimental data\cite{Zhang09,notazhang} (taken at 300K). At higher pressures
 there are no experimental data to compare with. It is clear how DFT-GGA
overestimate the distortion in the $ab$ plane (the $a/b$ ratio), the $c$ axis absolute value and the decrease
of the $c$ axis with increasing pressure (bottom panel).
On the opposite our calculations reproduce pretty well the $a$ axis absolute value and the reduction of the $a$ axis
 with increasing pressure.

\begin{figure}
\includegraphics[scale=0.5]{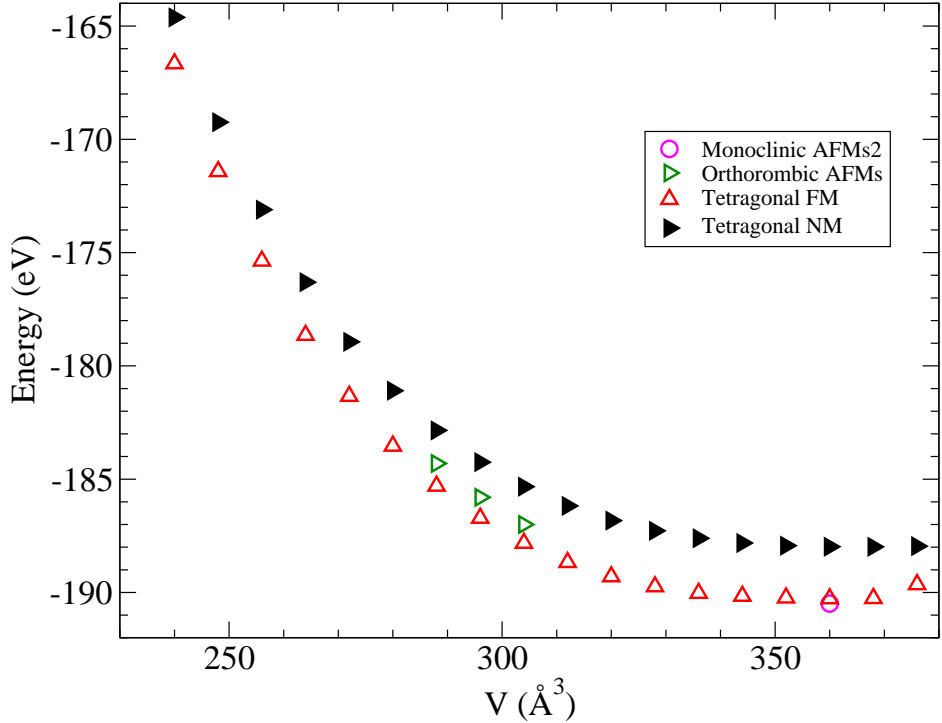}
\caption{(Color online). Total energy versus unit cell volume for a 17 Fe and 16 Te atoms cell.}
\label{fig:etot_17Fe}
\end{figure}

\begin{figure}
\includegraphics[scale=0.5]{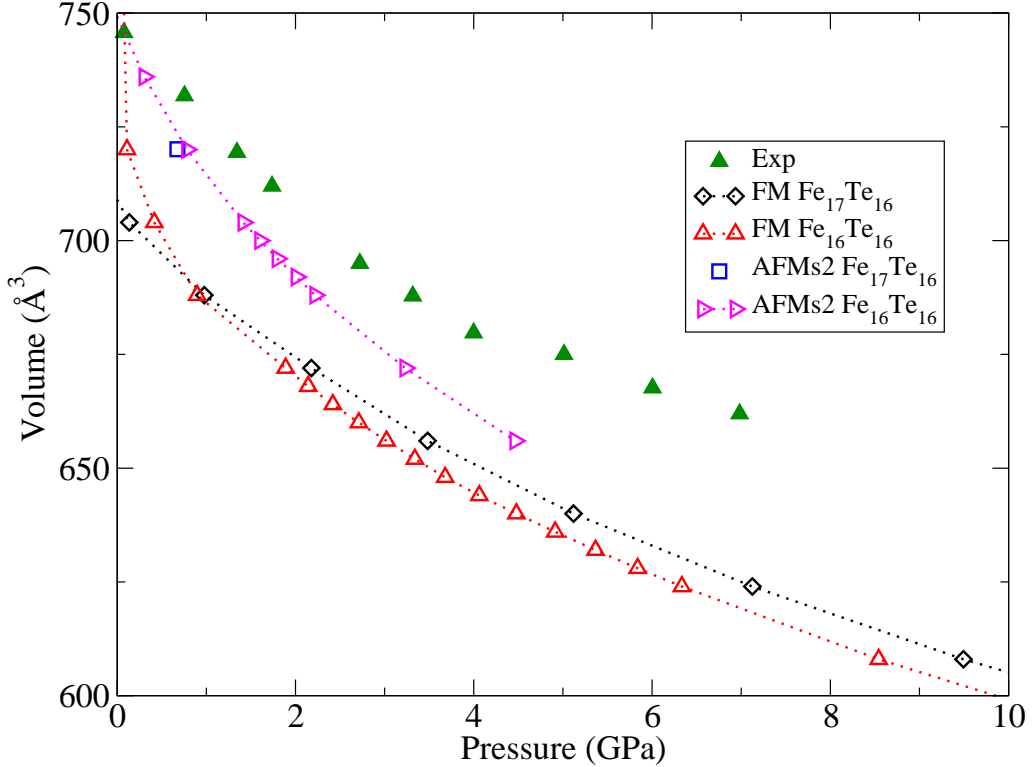}
\caption{(Color online) Cell volume as a function of pressure in Fe$_{17}$Te$_{16}$ and FeTe.}
\label{fig:Pi_Vi_17Fe}
\end{figure}

Although the qualitative agreement with experiments is good and reasonable, the quantitative comparison should be improved.
The inaccuracy of DFT in descring pnictides, is a possible explanation, but presence of excess Fe, which is always present in real samples (and not taken into account in most of simulations) could be a possible source of error.
 In fact, Ref.~\onlinecite{Zhang09} and Ref.~\onlinecite{Marina10} both estimates a 5\% concentration of excess Fe
in FeTe.
In order to clarify this point, always invoked to explain possible differences between theoretical predictions and experimental results,
  we performed representative calculations in a supercell with 17 Fe atoms and 16 Te atoms, which
 corresponds to Fe$_{1,0625}$Te. 
 The atoms where arranged in a $2a\times 2a\times 2c$ cell, and the excess Fe
occupies the so-called $2a$ site, which is roughly coplanar with Te atoms. 
Fig.\ref{fig:etot_17Fe} shows the total energy as a function of volume for representative magnetic orders. The excess Fe does not modify
substantially the energy sequence. Although possible disorder effects can be present this result 
poses stringent limitations on the effect of Fe excess.
At the same time, Fig.\ref{fig:Pi_Vi_17Fe} shows that the effect of excess Fe on structural properties is not
 crucial, being the compressibility with and without excess Fe nearly the same.

\begin{figure}
\includegraphics[scale=0.5]{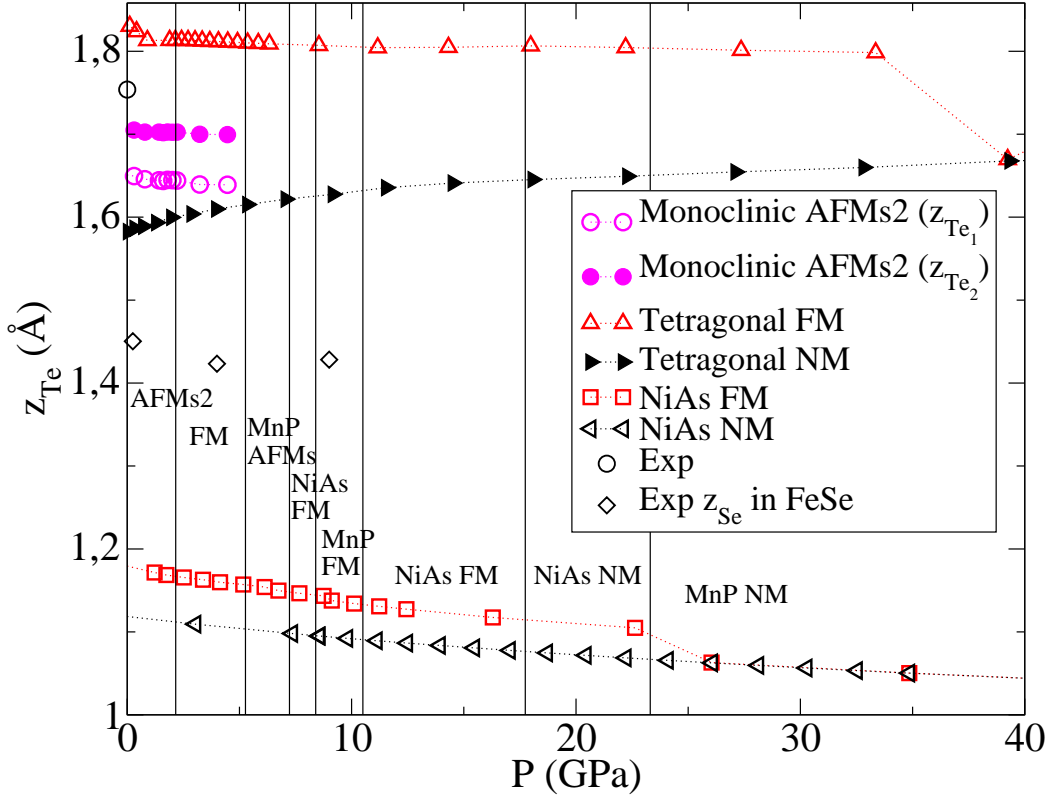}
\caption{(Color online). Te height above the Fe plane (in \AA) versus pressure. Experimental data from Ref.\onlinecite{Marina10}
 (black circle) and Ref.\onlinecite{Margadonna2009} (black rhombus).}
\label{fig:P_zTe}
\end{figure}

As discussed above there is a pressure range in which we obtain a ferromagnetic ordering along the Fe planes,
in agreement with the available experiments\cite{Bendele2012}, a fact which was not observed in
other pnictides based superconductors. However, the P based ones, LaCoPO is found to be ferromagnetic
 at ambient pressure\cite{Jin2009, Profeta2013}, and therefore not a superconductor, similarly to the case of FeTe.
The anion height above the Fe planes, is invoked as a possible parameter correlated with both the superconducting T$_c$ and the magnetic phase.\cite{kuroki2009,Moon2010,Mizuguchi2010} 
Thus, in order to better clarify the origin of the high pressure ferromagnetic phase, 
we report  in Fig.\ref{fig:P_zTe} the Te height over Fe planes. It is in fact know that
the Fe  moment depends significantly from this quantity. In the AFMs2 phase we have two Te sites with two
different heights.  The corresponding calculated average is smaller than the experimental value by 
about 0.1 \AA; in the FM phase, on the other hand, the theoretical value is larger by $\approx 0.5$ \AA~ and,
 finally, in the NM phase $z_{\rm Te}$ is $\approx 1.5$ \AA~ smaller than in experiment (which is a well known result).
It is clear that in FeTe the anion height above Fe planes is bigger than the Se height in FeSe and the As
height in LaFeAsO, which  can be easily understood considering that the covalent radius of Te atom is  bigger than Se and As.
In agreement with Ref.\onlinecite{Moon2010}, we find that high values of $z_{Te}$ favors FM alignment, confirming that anion height determines the magnetic order.  

The existence of ferromagnetic ordering is puzzling since it is not clear if it may be due to a deficiency
of GGA functional to describe properly the binding within the solid (notably a higher $c$ that experiment)
or this is a feature that distinguish this system from other compounds in the chalcogenide family.

\begin{figure}
\includegraphics[scale=0.5]{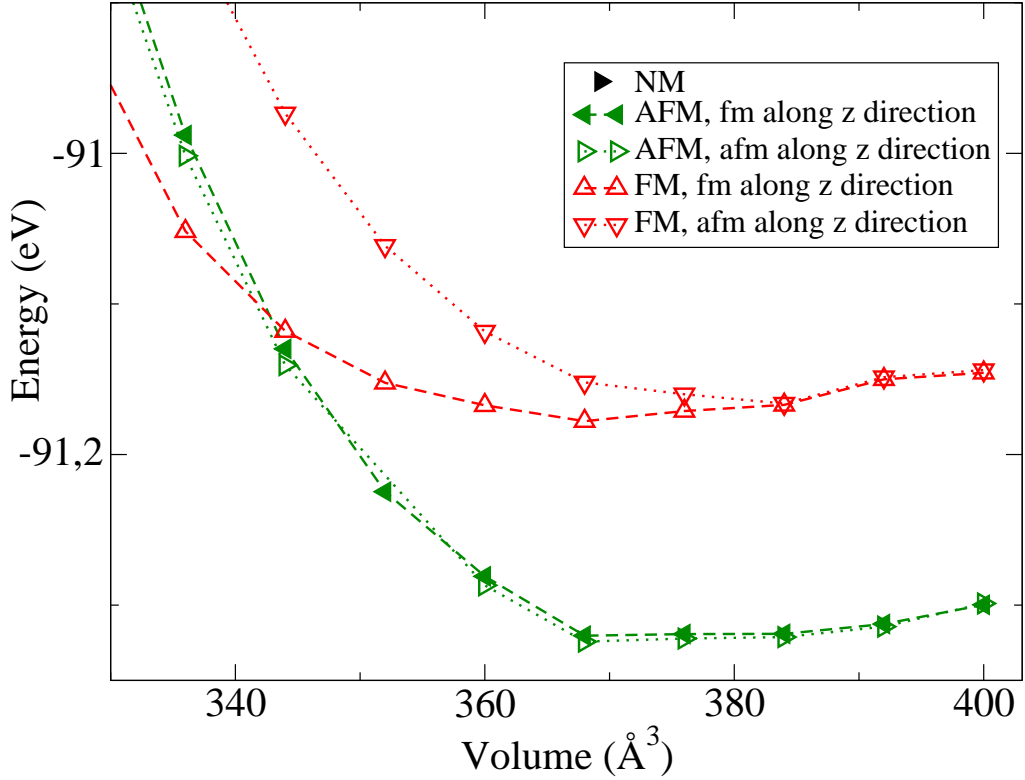}
\caption{(Color online). Total energy versus unit cell volume for an 8 Fe atoms cell doubled in the $z$ direction.
For each in plane magnetic ordering two different out of plane magnetic ordering were considered: ferromagnetic and antiferromagnetic.}
\label{fig:etot_zdir}
\end{figure}

\begin{figure}
\includegraphics[scale=0.5]{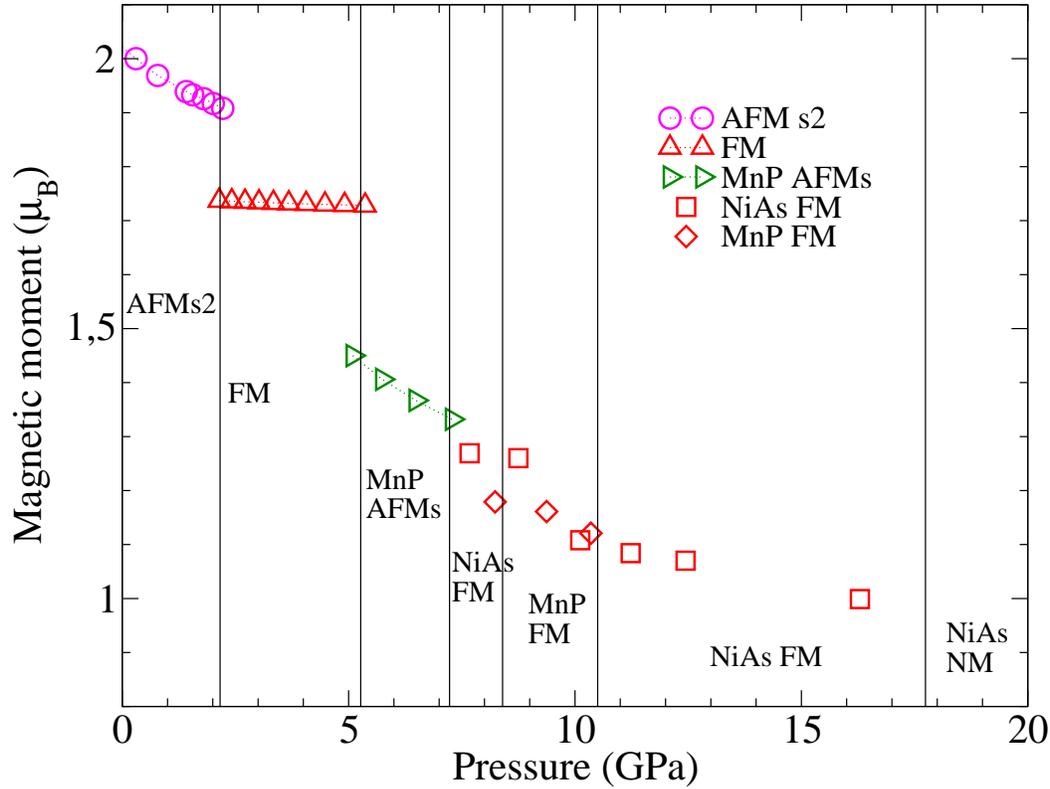}
\caption{(Color online). Magnetic moment on Fe site as a function of pressure.}
\label{fig:P_mu}
\end{figure}

As expected, the magnetic moment $\mu$ at Fe site decreases monotonically with increasing pressure
 (Fig.\ref{fig:P_mu}).
At low pressure, in the AFMs2 phase, the calculated magnetic moment is $\mu = 2\;\mu_B$, while experiments
report typical values around $2.5\;\mu_B$\cite{Marina10},$2.25\;\mu_B$\cite{Li2009}.
 In the MnP/NiAs pressure range of stability $\mu$ is nearly independent from the assumed structure.
 At $P\approx$17 GPa,
above which the system is non magnetic, $\mu \approx 1\;\mu_B$. Interestingly, the ferromagnetic
ordering in the PbO structure persists under compression till very high pressures. In fact, FM order
would survive up to pressures as big as $\approx$35 GPa, if it were the stable phase. 

Another interesting issue on pure FeTe is the lack of superconductivity up to $P=$19GPa\cite{Okada2009},
 the maximum pressure reached till now experimentally, this is the opposite to what happens in FeSe.\cite{Margadonna2009,Medvedev2009}
Our calculations reveal that the transition from magnetic ordering to non magnetic ordering with increasing pressure
 occurs at about 17GPa and involves a ferromagnetic ordered phase. So the strong fluctuations which likely manifest
near the transition are of ferromagnetic nature, oppositely to what happens in FeSe, and do not promote a
superconduting transition.

\section{Conclusions}
We present first-principle results on FeTe under hydrostatic pressure. We confirm the experimental evidence of 
an AFMs2 magnetic order at ambient pressure and predict several phase transitions
between magnetically ordered phases under pressure, until the magnetization goes to zero as 
the pressure increases above $P=$17GPa.
We find FM to be the ground state for pressure  between 2.1GPa and 17GPa. This finding
is in agreement with experimental results. We speculate that this may be the discriminant between the behavior of 
FeTe toward superconductivity with respect to other chalcogenides of the same family.  

\begin{acknowledgments}
We kindly acknowledge Prof. Liling Sun for sharing raw data.
We acknowledge computational support by CASPUR through Standard HPC Grant 2012 std12-162 and std12-000. M.M. acknowledges support from the Regione
Sardegna, through the felloship \textit{Studio computazionale dei superconduttori a base di ferro} within the program POR SARDEGNA FSE 2007-2013. F.B. acknowledges support by the FP7 European project SUPER-IRON (grant agreement No. 283204).
G.P. acknowledges supported by the FP7 European project SUPER-IRON (grant agreement No. 283204), 
by a CINECA- HPC ISCRA grant and by an HPC grant at CASPUR.
\end{acknowledgments}


\begin{thebibliography}{1}
\bibitem{notazhang} Absolute value of lattice constants were kindly provided by Liling Sun because they were not published in Ref.\onlinecite{Zhang09}.
\end{thebibliography}
\end{document}